\documentclass[letterpaper,10pt]{article} 

\usepackage{osameet3} 

\usepackage{amsmath,amssymb}
\usepackage[colorlinks=true,bookmarks=false,citecolor=blue,urlcolor=blue]{hyperref} 

\usepackage{comment}

\usepackage[font=small]{caption}
\usepackage[font=footnotesize]{subcaption}
\begin{document}
\title{High-Resolution Trans-Oceanic Distributed Acoustic Sensing Enabled by a Bi-Directional Sensor Implementation}


\vspace{-0.5cm}
\author{
    M.~Mazur\textsuperscript{(1)}, 
    N.~K.~Fontaine\textsuperscript{(1)}, 
    R.~Ryf\textsuperscript{(1)}, 
    M.~Karrenbach\textsuperscript{(2)}, 
    K.~M.~McBrian\textsuperscript{(3)}, 
    K.~L.~McLaughlin\textsuperscript{(3)}, 
    B.~J.~Sperry\textsuperscript{(3)}, 
    A.~G.~Butler\textsuperscript{(3)},
    V.~Kamalov\textsuperscript{(4)},  
    L.~Dallachiesa\textsuperscript{(1)}, \\
    E.~Burrows\textsuperscript{(1)},
    D.~Winter\textsuperscript{(1)},    
    H.~Chen\textsuperscript{(1)}, 
    J.~Naik\textsuperscript{(5)},
    K.~Padmaraju\textsuperscript{(5)},
    A.~Mistry\textsuperscript{(5)},    
    D.~T.~Neilson\textsuperscript{(1)}
}

\maketitle                  

\address{
    \textsuperscript{(1)} Nokia Bell Labs, 600 Mountain Ave., Murray Hill, NJ 07974, USA.
    \textcolor{blue} {\underline{mikael.mazur@nokia-bell-labs.com}} \\
    \textsuperscript{(2)} Seismics Unusual, LLC, Brea, CA 92821, USA \\
    \textsuperscript{(3)} Leidos Inc., Arlington, VA 22203, USA  \\
    \textsuperscript{(4)} Valey Kamalov LLC, Gainesville, FL 32607, USA\\
    \textsuperscript{(5)} Nokia Advanced Optics, 171 Madison Ave, New York City, NY 10016, USA \\
}
\copyrightyear{2026}
 
\vspace{-6mm}
\begin{abstract}   
We demonstrate continuous distributed acoustic sensing over a 4400\,km long undersea cable. Bi-directional operation improves the strain signal-to-noise rate by $>$20\,dB, enabling 88000 \\50-m-spaced measurement points at a nominal telecom launch power. 
\end{abstract} 
\vspace{-0mm}   
\section{Introduction}
\label{sec:intro}
\vspace{-2mm}
Deep ocean monitoring is currently limited to sparse coastal installations. To bridge this gap, interest has surged in repurposing undersea telecommunication cables for real-time seismic and environmental sensing~\cite{Bruce2019}. This global approach would enhance earthquake early warnings \cite{Allen2019} and address the majority of "grand challenges" in seismology \cite{Lay2009}. While Distributed Acoustic Sensing (DAS) has demonstrated potential for seismic and oceanic monitoring~\cite{Williams2019}, it has traditionally been limited to the first and last spans of undersea cables. We recently overcame this limitation by using a long-reach OFDR prototype, making the entire cable length accessible for sensing with sub-repeater-spacing resolution \cite{Mazur2025_ECOC}. Compared to other methods such as per-span interferometry~\cite{Marra2022}, it enables a 1000x improvement in spatial resolution. However, global scaling requires addressing two key hurdles: sensing power levels must be minimized to preserve telecom throughput, and the signal-to-noise ratio (SNR) must be improved to overcome the $\sim$40 dB loss typical of loopback couplers. 

We address these challenges by demonstrating distributed strain sensing with 50-m spatial resolution over a 4,400-km undersea cable featuring about 100 submerged optical repeaters. The cable connects California to Hawaii, supporting approximately 20 channels on a 100-GHz grid. By employing two separate long-reach sensing systems, enabled by wavelength multiplexing (WDM) of narrow-band signals, we simultaneously monitor the cable in both directions at a nominal launch power of -3\,dBm for a 100-GHz channel slot. The bi-directional implementation allows for overcoming SNR limitations, providing $>$20\,dB improvement along the entire cable. This allows for distributed sensing with 50-m spatial resolution with a noise floor of $\approx$1\,n$\epsilon$/s/$\sqrt{Hz}$ at 0.2\,Hz along the entire cable. We furthermore verify the performance by a detailed analysis of the 30-11-2025 M4.9 earthquake in Susanville California, about 1250\,km from the cable landing station and high-resolution recording of seismic waves on the deep ocean floor. Our results demonstrate that continuous, transoceanic, sensing with sub-100-m resolution is achievable using only the power and bandwidth of a single telecom channel. This paves the way for integrating long-reach sensing into all undersea infrastructure, transforming the global subsea network into a planetary-scale instrument for real-time Earth monitoring.

\begin{figure*}[hb]
   \centering
    \includegraphics[width=1\linewidth]{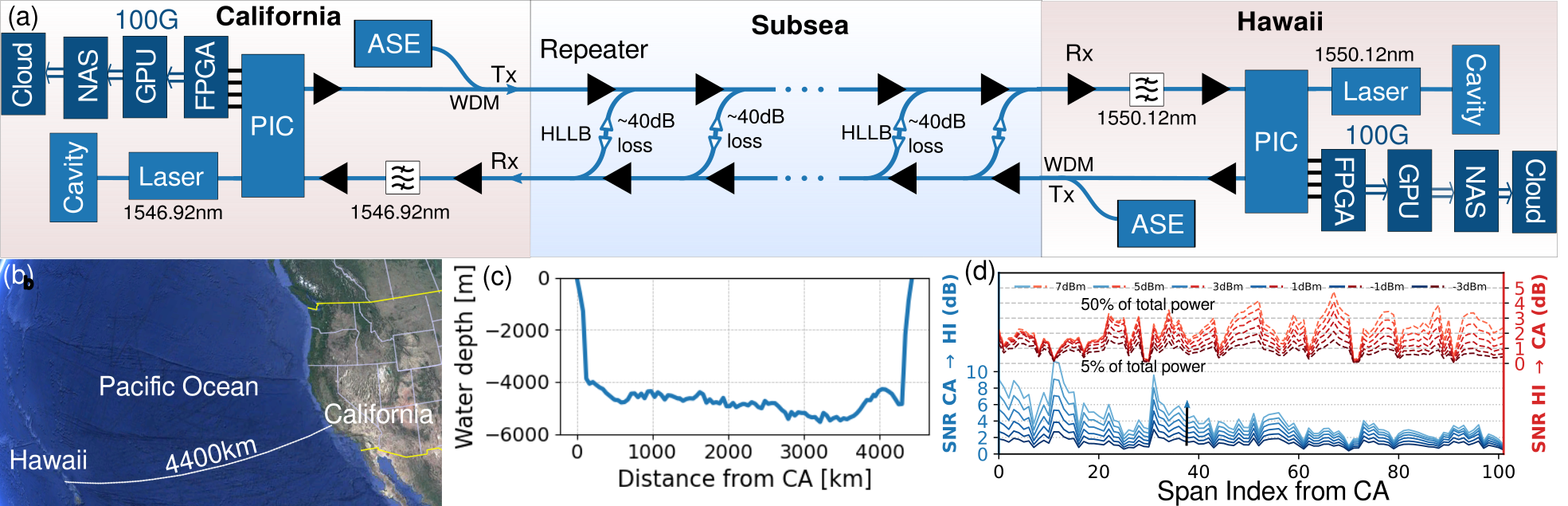}
    \caption{(a) Experimental setup for the bi-directional implementation. The sensing signals were combined with ASE loading using 100-GHz WDM filters.(b) Map showing the 4400-km undersea cable connecting California to Hawaii, USA. (c) Cable depth profile. More than 4000\,km is at depths exceeding 4000\,m. (d) Estimate of the optical signal-to-noise-ratio for various launch power from both directions. Lowest power of -3\,dBm equals allocating one 100-GHz WDM channel for sensing. Correspondingly, 7\,dBm equals allocating half the available power for sensing.  A launch power of -3dBm was used in this experiment. 
    \label{fig:setup}}\vspace{-2mm}
\end{figure*}
\section{Experimental Setup}\vspace{-1mm}
The experimental setup for the long-reach DAS system is shown in Fig.~\ref{fig:setup}(a). It uses a bi-directional configuration with two long-reach OFDR systems~\cite{Mazur2025_ECOC}. Each system consists of a cavity-stabilized fiber laser, which is externally fed to a real-time sensing unit in a homodyne configuration. The sensing unit use a dual-polarization transmit and receive photonic integrated circuit (PIC), connected to an FPGA (AMD/Xilinx ZU48DR). The FPGA generates and detects dual-polarization chirped pulses with 250\,MHz bandwidth and 67\,ms pulse duration. The real-time receiver processing is split between the FPGA and GPU, which are connected via a 100-Gbit/s optical link. The system outputs all intensity, phase and polarization information in parallel, supporting a wide range of uses cases from environmental sensing to repeater monitoring. The phase-sensitive "DAS" data outputting phase change, or strain rate, was calculated using standard DAS processing very similar to ref~\cite{waagaard2021real}, with a tunable spatial resolution down to about 10\,m. 
\begin{figure*}[ht!]
   \centering
    \includegraphics[width=1\linewidth]{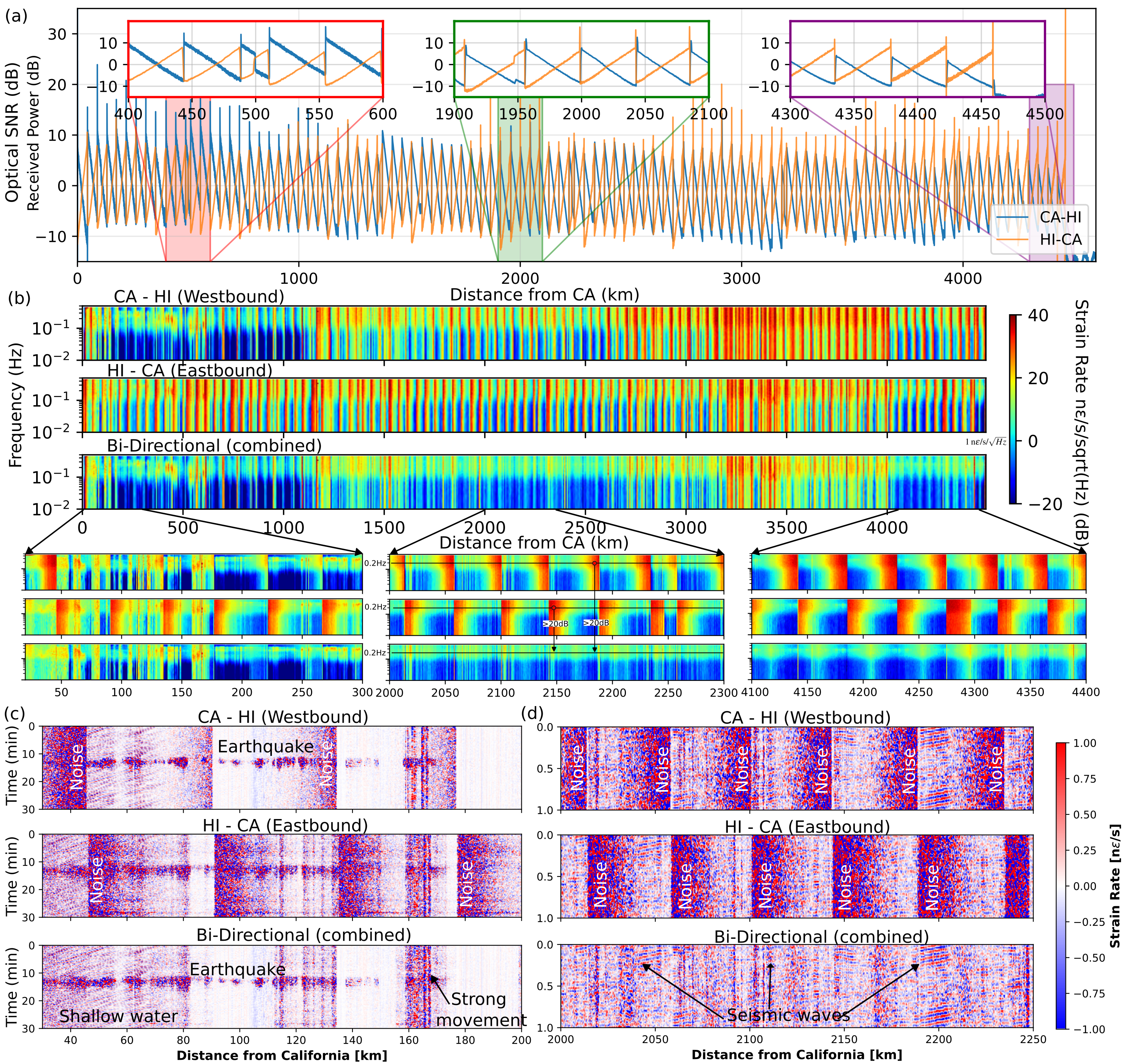}
    \caption{(a) Received power, or corresponding SNR, from the bi-directional sensing system. The lowest SNR point is shifted to the center of each span. (b) Individual spectrograms for each direction and from the combined signals. Bi-directional operation clearly reduces the noise level, improving all spans with $>$20\,dB at 100\,mHz and up to 40\,dB at 10\,mHz. The remaining red stripes are from cable movements and not limited by noise. (c) Corresponding waterfall plot for the first 200\,km, showing high SNR detection of a M4.9 earthquake about 1250\,km from the cable shore. (d) Waterfall diagram for a 1-minute trace with 50-m spatial resolution 2000\,km offshore California. Combining both signals overcomes the noise limitation, enabling clear tracking of seismic waves on the ocean floor, which can be used for tomographic studies of earth's structure.
    \label{fig:results}}\vspace{-6mm}
\end{figure*}

The systems were installed on a $\approx$4400\,km long undersea cable connecting California and Hawaii, as shown in Fig.~\ref{fig:setup}(b). It has about 100 submerged repeaters with a median span length around 45\,km. The cable depth profile is shown in Fig.~\ref{fig:setup}(c). Most of the cable is at a depth exceeding 4000\,m. High-loss loopback couplers (HLLBs)~\cite{Otani} with about 40\,dB added loss are present in each repeater, enabling backscattered light to bypass the inline isolators, as conceptually shown in Fig.~\ref{fig:setup}(a). In addition, two equivalent HLLBs are inserted on the shore sides, enabling backscattered light from the first spans to be coupled into the Rx fiber. The data is stored locally and synchronized using a combination of referenced system clocks and pulse-level time-counting. 
We rely on a single fiber-pair implementation. The backscattered light from the CA to HI probe therefore co-propagates with HI to CA probe. 
To avoid interference, the two signals must be offset in wavelength. However, given the sub-GHz total signal bandwidth, both easily fit within a standard 50 or 100\,GHz wavelength slot. 
However, due to laser availability limitations, the CA-HI system used an edge channel at 1546.92\,nm and the HI-CA system used 1550.12\,nm. 
The usable cable bandwidth was about 2\,THz and the OFDR channel launch power was -3\,dBm. The rest of the spectrum was filled with noise at nominal power density.\vspace{-2mm}

\section{Results}\vspace{-2mm}
Figure~\ref{fig:setup}(d) shows optical SNR scans from the bi-directional implementation. 
The allocated power is varied between -3\,dBm and +7\,dBm. Equivalently, -3\,dBm is the nominal launch power for the allocated 100-GHz channel. 
In contrast, +7\,dBm corresponds to allocating 50\% of the total available launch power to the sensing signal. 
We can observe $\approx$5\,dB improvement when increasing the launch power to 7\,dBm. These results are also in line with previous lab experiments~\cite{Ryf2026}. In addition, we observe a larger roll-off in SNR for the CA to HI direction, which is due to 1546.92\,nm being an edge channel with worse repeater performance. The fiber loss is about 0.2\,dB/km (0.4\,dB/km for backscattered light) and 5\,dB therefore corresponds to 12.5\,km additional reach. However, we here focus on high resolution sensing with minimal impact on telecom channels and therefore use -3\,dBm launch power and the bi-directional approach to cover the entire cable instead. 

The bi-directional advantage is clear in Fig.~\ref{fig:results}(a), showing the combined measured distributed intensity profile. We observe that the optical SNR crosses zero around half-way through the span, matching the results in Fig.~\ref{fig:setup}(d). 
The measured strain rate power spectral density  is shown in Fig.~\ref{fig:results}(b), showing a distinct noise level improvement along the entire cable for the bi-directional approach. Zoom-ins focusing on shore-ends together with the middle point highlights the achieved improvement. 
Combining both signals, we achieve a typical noise level of 1\,n$\epsilon$/s/$\sqrt{Hz}$ at 0.2\,Hz, a reduction of $>$20\,dB compared to the single-ended approach. At 0.01\,Hz, the typical improvement is close to 40\,dB. The residual "red" regions, are the results of increased background noise level. For example, up until about 80\,km the cable is at a water depth of $<100$\,m and around 3400\,km the cable is traversing rocky underwater mountain train with poor coupling to the ocean floor. 
A waterfall plot for the first 200\,km is shown in Fig.~\ref{fig:results}(c). Here we clearly see the arrival of surface waves from the 30-11-2025 M4.9 earthquake with epicenter in Susanville California, about 1250\,km from the cable landing station. We note two things, first, in line with Fig.~\ref{fig:results}(b), the bi-directional approach enables high SNR sensing along the entirety of each span, despite the fact that it is after $>$4000\,km propagation from the HI-CA system, the performance is still orders of magnitude better than for the CA-HI system alone. Secondly, we observe the varying cable coupling and background noise level from Fig.~\ref{fig:results}(b). In addition, we observe some point of strong motion such as around 165\,km. Figure~\ref{fig:results}(d) shows a waterfall representation of a 1-minute duration from Fig.~\ref{fig:results}(b). Here we clearly see the low-SNR regimes from both directions, together with the stitching improvement. Following stitching, we can clearly resolve seismic coherent waves propagating on the ocean floor. These are waves in the secondary microseismic bands, originating from nonlinear pressure interaction between opposing trains of ocean waves. Monitoring and tracking of these waves can be used to perform ambient noise tomography to understand earth's crust and layers underneath the ocean.\vspace{-3mm}

\section{Conclusion}\vspace{-2mm}
We demonstrated distributed acoustic sensing with 50-m resolution over a 4400-km undersea cable, effectively creating an array of 88,000 sensors on the ocean floor. By employing a bi-directional architecture, we achieved a noise floor of $\approx$1\,n$\epsilon$/s/$\sqrt{Hz}$ at 0.2\,Hz across the entire span. This approach successfully compensates for the $\approx$40\,dB attenuation of the loop-back couplers, delivering high SNR monitoring while consuming the optical power of only a single 100-GHz telecom channel. Our results confirm the viability of high-resolution sensing along transoceanic cables with negligible impact on the data transmission capacity. This paves the way for transforming the global subsea network into a planetary-scale instrument, illuminating the unmonitored deep ocean to revolutionize seismic monitoring and tsunami early warning systems.






\vspace{-4mm}

\end{document}